\documentclass{LMCS}
\usepackage{amssymb}
\usepackage{proof,pb-diagram}
\usepackage{hyperref}

\newtheorem{mytheorem}{Theorem}[section]
\newtheorem{mydefinition}[mytheorem]{Definition}
\newtheorem{myproposition}[mytheorem]{Proposition}
\newtheorem{mylemma}[mytheorem]{Lemma}
\newtheorem{myremark}[mytheorem]{\it Remark}
\newtheorem{mycorollary}[mytheorem]{Corollary}
\newtheorem{myexample}[mytheorem]{\it Example}
\newenvironment{theorem}{\begin{mytheorem} }{\end{mytheorem}}
\newenvironment{definition}{\begin{mydefinition} \em}{\end{mydefinition}}
\newenvironment{proposition}{\begin{myproposition} }{\end{myproposition}}

\newenvironment{corollary}{\begin{mycorollary} \em}{\end{mycorollary}}

\newcommand{\sflet}{\mbox{\sf let}}
\newcommand{\sfbe}{\mbox{\sf be}}
\newcommand{\sfin}{\mbox{\sf in}}
\newcommand{\letbe}[3]{\sflet~#1~\sfbe~#2~\sfin~#3}
\newcommand{\cps}[1]{[\![#1]\!]}
\newcommand{\pair}[1]{\langle #1 \rangle}
\newcommand{\calC}{\mathcal{C}}

\newcommand{\name}[1]{\mbox{[}#1\mbox{]}}
\newcommand{\mymu}{\mbox{\boldmath$\mu$}}
\newcommand{\myname}[1]{\mbox{\bf [}#1\mbox{\bf ]}}

\newcommand{\comment}[1]{}

\def\today{\number\day\space
\ifcase\month\or
January\or February\or March\or April\or May\or June\or
July\or August\or September\or October\or November\or December\fi
\space\number\year}

\def\doi{2 (3:3) 2006}
\lmcsheading%
{\doi}
{1--22}
{}
{}
{Dec.~16, 2005}
{Jul.~27, 2006}
{}

\begin{document}

\title[Relational Parametricity and Control]{Relational Parametricity and Control}
\author{Masahito Hasegawa}
\address{Research Institute for Mathematical Sciences, 
                    Kyoto University, 
                    Kyoto 606-8502 Japan,
        and
        PRESTO, Japan Science and Technology Agency}
\email{hassei@kurims.kyoto-u.ac.jp}


\begin{abstract}
We study the equational theory of Parigot's second-order
$\lambda\mu$-calculus in connection with a call-by-name
continuation-passing style (CPS) translation
into a fragment of the second-order $\lambda$-calculus.
It is observed that the relational parametricity on
the target calculus induces a
natural notion of equivalence on the $\lambda\mu$-terms. 
On the other hand, the unconstrained relational parametricity
on the $\lambda\mu$-calculus turns out to be inconsistent.
Following these facts, we propose to formulate the 
relational parametricity on the $\lambda\mu$-calculus in 
a constrained way, which might be called ``focal parametricity''.\\
\mbox{}\\
\mbox{}\\
{\em Dedicated to Prof. Gordon Plotkin on the occasion of his sixtieth birthday}
\end{abstract}

\maketitle

\section{Introduction}

The {\em $\lambda\mu$-calculus},
introduced by Parigot \cite{Par92}, has been
one of the representative term calculi for
classical natural deduction, and widely studied
from various aspects.
Although it still is an active research subject, 
it can be said that we have some
reasonable understanding of the first-order propositional
$\lambda\mu$-calculus:
we have good reduction theories, well-established CPS semantics and
the corresponding operational semantics, and also some canonical 
equational theories enjoying semantic completeness 
\cite{HS02,Ong96,OS97,Sel01,SR98}. 
The last point cannot be overlooked, as such complete axiomatizations provide 
deep understanding of equivalences between proofs and also of the 
semantic structure behind the syntactic presentation.

The {\em second-order $\lambda\mu$-calculus} ($\lambda\mu2$), again due to 
Parigot \cite{Par97}, has been studied in depth as a calculus for  
second-order classical natural deduction. In particular, 
strong normalization results of $\lambda\mu2$ \cite{Par97,NT03} and its 
extensions, e.g. with inductive types \cite{Mat01}, 
have been a central research topic,
because of the proof-theoretical importance of strong normalization.
However, for $\lambda\mu2$,
it seems that there are few attempts of giving an equational theory 
supported by some fine semantic structure. This situation is
rather frustrating, since without such equational and semantic accounts,
we cannot discuss e.g. the correctness of the impredicative encoding
of the datatypes in $\lambda\mu2$. 
For the second-order $\lambda$-calculus $\lambda2$ (system F) 
\cite{Gir72,Rey74}, a subsystem of
$\lambda\mu2$, there are several beautiful results on the {\em relational
parametricity} \cite{Rey83} and the universal properties of 
impredicative constructions \cite{ACC93,Ryu94,PA93,Tak98,Wad89},
e.g. that $\mu X.\sigma = \forall X.(\sigma\rightarrow X)\rightarrow X$
(where $\sigma$ covariant in $X$) gives an initial algebra of
the functor $\Lambda X.\sigma$ in a suitable sense. 
We certainly wish to have such a story for $\lambda\mu2$ too.

This work is an attempt to identify such an equational theory which is
backed up by certain semantic structures. Specifically, we propose
a relational parametricity principle which is sound and sufficiently
powerful for deriving such equivalences on the $\lambda\mu$-terms.

\subsection{Parametric CPS semantics}

We first consider the semantics of $\lambda\mu2$
given by a CPS-translation into a fragment of $\lambda2$
--- that of the second-order existential types $\exists X.\tau$, 
conjunction types $\tau_1\wedge\tau_2$,
and arrow types $\tau\rightarrow R$ 
into a distinguished type $R$ (this choice of the
target calculus is due to a recent work of Fujita \cite{Fuj04}).
The translation $(-)^\circ$ sends a type variable $X$ to 
$X$, arrow type $\sigma_1\rightarrow\sigma_2$ to 
$(\sigma_1^\circ\rightarrow R)\wedge\sigma_2^\circ$,
and the universal type $\forall X.\sigma$ to $\exists X.\sigma^\circ$ --- 
while a term $M:\sigma$ is sent to $[\![M]\!]:\sigma^\circ\rightarrow R$.
It can be
considered as a natural extension of Streicher's call-by-name CPS
translation \cite{Sel01,SR98,Thi04}.
It follows that this translation 
already gives a reasonable equational theory on 
$\lambda\mu2$, in that it validates the standard $\beta\eta$-equalities%
\footnote{We can say more -- we can show that this
CPS-semantics is sound and complete with respect to the 
$\beta\eta$-theory of $\lambda\mu2$. This result, together with
further syntactic analysis of this CPS translation, will appear
in a forthcoming paper with Ken-etsu Fujita.}.
In fact, this is a consequence of a fibred version of the
"category of continuations" construction \cite{HS02,Sel01,SR98}. 

However, this is just a starting point; we observe that, if some of the
impredicative
constructions in the target calculus satisfy certain universal properties
(e.g. $\exists X.X$ is a terminal object)
which follow from the relational parametricity, then
so do the impredicative constructions in the source $\lambda\mu2$-calculus
--- but not quite in the way that we first might expect. For instance,
the type $\bot=\forall X.X$ does {\em not}
give an initial object (cf. \cite{Sel03})
---
instead it plays the role of the falsity type 
(or the ``answer type'');
in fact, we have a {\em double-negation elimination} from 
$(\sigma\rightarrow\bot)\rightarrow\bot$ to $\sigma$ for any $\sigma$
which actually is an algebra of the double-negation monad
$((-)\rightarrow\bot)\rightarrow\bot$.
As another major example, $\forall X.(\sigma\rightarrow X)\rightarrow X$
does {\em not} give an initial algebra of $\Lambda X.\sigma$; it gives 
an initial algebra of $\Lambda X.(\sigma\rightarrow\bot)\rightarrow\bot$
--- not with respect to all terms but to a certain class of terms
(the ``focal terms'', to be mentioned below).
In particular, if $X$ is not free in $\sigma$, 
$\forall X.(\sigma\rightarrow X)\rightarrow X$ is isomorphic not to 
$\sigma$ but to $(\sigma\rightarrow\bot)\rightarrow\bot$.
In short, impredicative encodings in $\lambda\mu2$ get extra double
negations, and the relational parametricity of $\lambda2$ is not consistent
with the equational theory of $\lambda\mu2$ induced by the CPS semantics.
As a consequence, we cannot encode cartesian products
in $\lambda\mu2$, though they can be added easily. Also
we cannot express the classical disjunctions \cite{Sel01}, 
though they can be 
added without changing the target of the CPS translation.

\subsection{Focal parametricity}

These results suggest that the CPS translation into
parametric target calculus gives a reasonable 
semantic foundation and equational theory for $\lambda\mu2$,
which is sufficient for obtaining various interesting results.
However, here the parametricity is used rather indirectly, via the
CPS translation; 
we also wish to have a decent notion of 
parametricity directly within $\lambda\mu2$.
To figure out what sort of parametricity principle can be expected
for $\lambda\mu2$, recall the following fact on $\lambda2$ with 
parametricity: given a polymorphic term 
$M:\forall X.F[X]\rightarrow G[X]$ (with $X$ covariant in
$F$ and $G$) and types $\sigma_1$, $\sigma_2$,
the instances
$M\,\sigma_1:F[\sigma_1]\rightarrow G[\sigma_1]$ and
$M\,\sigma_2:F[\sigma_2]\rightarrow G[\sigma_2]$ 
obey the {\em naturality}, in that the following diagram
$$
\begin{diagram}
\node{F[\sigma_1]}
 \arrow{s,l}{M\,{\sigma_1}}
 \arrow{e,t}{F[f]}
\node{F[\sigma_2]}
 \arrow{s,r}{M\,\sigma_2}\\
\node{G[\sigma_1]}
 \arrow{e,b}{G[f]}
\node{G[\sigma_2]}
\end{diagram}
$$
commutes for {\em any} $f:\sigma_1\rightarrow\sigma_2$.
This is no longer true for $\lambda\mu2$. For example, 
let $F[\sigma]=(\sigma\rightarrow\bot)\rightarrow\bot$,
$G[\sigma]=\sigma$ and $M$ be the double-negation elimination
(which does not exist in $\lambda2$); 
then the naturality for arbitrary maps 
implies inconsistency 
--- we get $\sigma\simeq(\sigma\rightarrow\bot)\rightarrow\bot$ 
for every $\sigma$ by letting $f$ be the obvious map from
$\sigma$ to $(\sigma\rightarrow\bot)\rightarrow\bot$, which 
is enough to kill the theory \cite{LS86}. 
Similar result can be observed for other ``classical'' proofs,
e.g. of the Peirce law.

To this end, we look at the {\em focus} \cite{Sel01}
({\em centre} \cite{PR97,Thi97}, {\em C-maps} \cite{HS02})
of $\lambda\mu2$; a focal map is 
no other than an algebra morphism 
between the the 
double-negation monad mentioned above, i.e., a map making the naturality
diagram for the double-negation elimination commute. 
It follows
that a notion of relational parametricity on $\lambda\mu2$
in which the construction of the graph relations is allowed only for 
focal maps is consistent, as there are nontrivial models. 
Together with the definability (fullness) of the CPS translation, 
we see that it is 
at least as powerful as the parametricity on the CPS target calculus
which we have mentioned above,
thus gives a powerful principle for deriving the equivalences 
of terms in $\lambda\mu2$. 
(We actually conjecture that these two notions of parametricity do
agree, but it is open as of writing this article.)
This principle, which we shall call {\em focal parametricity}, 
should be a natural notion of parametricity for $\lambda\mu2$.
We will sketch some use of focal parametricity for deriving
``free theorems'' for $\lambda\mu2$ syntactically.

\subsection{Towards parametricity for computational effects}

At the conceptual and abstract level, this story closely resembles to
the study of {\em linear parametricity} and recursion \cite{BPR00,Plo93}.
In the case of linear parametricity, the graph relations are allowed to be
constructed only from the linear maps, and a linear map is 
an algebra map w.r.t. the lifting monad. 
We claim that, just like the linear parametricity gives a
solution of accommodating non-termination and recursion in 
the polymorphic setting (as advocated by Plotkin \cite{Plo93}), the 
focal parametricity provides a way of accommodating control features in 
the polymorphic setting. In short:

$$
\frac{\mbox{linear parametricity}}{\mbox{non-termination}}
=
\frac{\mbox{focal parametricity}}{\mbox{first-class control}}
$$

\mbox{}

As future work,
it would be an interesting challenge to  
find a unifying framework of linear parametricity and 
focal parametricity; it should be useful to have parametric polymorphism,
recursion, and control at once, as in the realistic programming languages
(cf. \cite{HK02,Kak02,KH03}). More ambitiously, we are keen to
see an adequate notion of parametricity for fairly general 
``effectful'' settings.
Possible starting points for this direction might include 
the ``parametricity graphs'' approach \cite{Dun02} which 
allows us to deal with parametricity at a general level
(including the linear parametricity as an instance),  
and
the ``category of linear continuations construction'' 
\cite{Has04}
which induces both the CPS translation and Girard translation as
special cases. 
See Section 7 for further discussions related to this issue.

\subsection{Construction of this paper}

The rest of this paper is organised as follows.
In section 2 and 3 we introduce the calculi and CPS-translation
which are the subject of this study. In section 4 we consider 
the implications of the relational parametricity on the 
CPS-target calculus. The focal parametricity is introduced in
section 5, followed by examples in section 6, including
focally initial algebras and the type of Church numerals. Section 7 
gives an alternative characterisation of focus, which 
suggests a generalisation of this work to a theory of parametricity
for general effects. 
We then give some concluding remarks in Section 8.

\section{The calculi}

\subsection{The second-order \texorpdfstring{\boldmath$\lambda\mu$}-calculus} 

The second-order $\lambda\mu$-calculus, $\lambda\mu2$, is 
given as follows. We essentially follow Parigot's formulation
\cite{Par97} (with some flavour from Selinger's \cite{Sel01}).
The types are the same as those of the second-order 
$\lambda$-calculus $\lambda2$:
$$\sigma~::=~X~|~\sigma\rightarrow\sigma~|~\forall X.\sigma$$
In a typing judgement $\Gamma\vdash M:\sigma~|~\Delta$,
$\Gamma$ stands for the typing context of variables,
while $\Delta$ for the context of names (continuation variables).

$$
\infer{\Gamma,x:\sigma,\Gamma'\vdash x:\sigma~|~\Delta}{}
$$

$$
\infer{\Gamma\vdash\lambda x^{\sigma_1}.M:\sigma_1\rightarrow\sigma_2~|~\Delta}
      {\Gamma,x:\sigma_1\vdash M:\sigma_2~|~\Delta}
$$

$$
\infer{\Gamma\vdash M\,N:\sigma_2~|~\Delta}
      {\Gamma\vdash M:\sigma_1\rightarrow\sigma_2~|~\Delta &
       \Gamma\vdash N:\sigma_1~|~\Delta}
$$

$$
\infer{\Gamma\vdash \Lambda X.M:\forall X.\sigma~|~\Delta}
      {\Gamma\vdash M:\sigma~|~\Delta & (X\not\in FTV(\Gamma,\Delta))}
$$

$$
\infer{\Gamma\vdash M\,\sigma_2:\sigma_1[\sigma_2/X]~|~\Delta}
      {\Gamma\vdash M:\forall X.\sigma_1~|~\Delta}
$$

$$
\infer{\Gamma\vdash \mu\alpha^{\sigma_1}.\name{\beta}M:\sigma_1~|~\Delta}
      {\Gamma\vdash M:\sigma_2~|~\alpha:\sigma_1,\Delta 
       & (\beta:\sigma_2\in\alpha:\sigma_1,\Delta)
      }
$$
The axioms for the equational theory are again the standard ones ---
note that we consider the extensional theory, i.e. with the $\eta$-axioms.
$$
\begin{array}{rcl}
(\lambda x^\sigma.M)\,N &=& M[N/x]\\
\lambda x^\sigma.M\,x &=& M ~~~(x\not\in FV(M))\\
(\Lambda X.M)\,\sigma &=& M[\sigma/X]\\
\Lambda X.M\,X &=& M ~~~(X\not\in FTV(M))\\
\\
\mu\alpha.\name{\beta}(\mu\gamma.M)
&=&
\mu\alpha.M[\beta/\gamma]\\
\mu\alpha^\sigma.\name{\alpha}M 
&=& M ~~~(\alpha\not\in FN(M))\\
(\mu\alpha^{\sigma_1\rightarrow\sigma_2}.M)\,N
&=&
\mu\beta^{\sigma_2}.M[\name{\beta}(-\,N)/\name{\alpha}(-)]\\
(\mu\alpha^{\forall X.\sigma_1}.M)\,\sigma_2
&=&
\mu\beta^{\sigma_1[\sigma_2/X]}.M[\name{\beta}(-\,\sigma_2)/\name{\alpha}(-)]\\
\end{array}
$$
In the last two axioms, we make uses of so-called "mixed substitution''; 
for instance,
$M[\name{\beta}(-\,N)/\name{\alpha}(-)]$ means replacing
occurances of the form $\name{\alpha}L$ in $M$ by $\name{\beta}(L\,N)$
recursively. 

In the sequel, we frequently use the following syntactic sugar.
First, we let $\bot$ be the type $\forall X.X$ --- the type of falsity.
We may also write $\neg\sigma$ for $\sigma\rightarrow\bot$.
Using $\bot$, we define the ``named term'' 
$$
\myname{\beta}M~\equiv~\mu\alpha^\bot.\name{\beta}M:\bot 
$$
(where $M:\sigma$, $\beta:\sigma$, with $\alpha$ fresh)
and the 
$\mu$-abstraction 
$$\mymu\alpha^\sigma.M:\sigma~\equiv~\mu\alpha^\sigma.\name{\alpha}(M\,\sigma)$$
for $M:\bot$.
It follows that 
$\mymu\alpha^\sigma.\myname{\beta}M=\mu\alpha^\sigma.\name{\beta}M$
holds. 

With this $\bot$, we can express 
the double-negation elimination in $\lambda\mu2$ by
making use of both the polymorphic and classical
features:
$$
C_\sigma=
 \lambda m^{\neg\neg\sigma}.\mymu\alpha^\sigma.m\,(\lambda x^\sigma.\myname{\alpha}x)
:\neg\neg\sigma\rightarrow\sigma
$$
As expected, we have
$C_\sigma\,(\lambda k^{\neg\sigma}.k\,M)=M$. 
The properties of $\bot$ and $C_\sigma$ 
will be further studied later under parametricity assumptions.

\subsection{Target: the \texorpdfstring{\boldmath$\{\exists,\wedge,\neg\}$}-calculus}

In tthe literature, the second-order $\lambda$-calculus ($\lambda2$)
is often taken as the target of the CPS translation for $\lambda\mu2$.
Fujita observed that it actually suffices to consider a fragment of
$\lambda2$ with negations, conjunctions and existential types
as a target \cite{Fuj04}. In this paper we follow this insight.
$$
\begin{array}{lcl}
\tau &::=& X~|~R~|~\neg\tau~|~\tau\wedge\tau~|~\exists X.\tau\\
\end{array}
$$
$\neg\tau$ can be considered as a shorthand of $\tau\rightarrow R$.
The type $R$ can be replaced by $\exists X.\neg X\wedge X$, but for 
simplicity we keep $R$ as a type constant.
The syntax of terms is a fairly standard one, though for
conjunctions we employ a slightly less familiar elimination rule
(with $\mathsf{let}$-binding) so that it parallels that of the
existential types.
$$
\infer{\Gamma,x:\tau,\Gamma'\vdash x:\tau}{}
$$

$$
\infer{\Gamma\vdash\lambda x^\tau.M:\neg\tau}
      {\Gamma,x:\tau\vdash M:R}
$$

$$
\infer{\Gamma\vdash M\,N:R}
      {\Gamma\vdash M:\neg\tau & \Gamma\vdash N:\tau}
$$

$$
\infer{\Gamma\vdash\pair{M,N}:\tau_1\wedge\tau_2}
      {\Gamma\vdash M:\tau_1 & \Gamma\vdash N:\tau_2}
$$

$$
\infer{\Gamma\vdash\letbe{\pair{x^{\tau_1},y^{\tau_2}}}{M}{N}:\tau_3}
      {\Gamma\vdash M:\tau_1\wedge\tau_2 &
       \Gamma,x:\tau_1,y:\tau_2\vdash N:\tau_3}
$$

$$
\infer{\Gamma\vdash\pair{\tau_2,M}:\exists X.\tau_1}
      {\Gamma\vdash M:\tau_1[\tau_2/X]}
$$

$$
\infer{\Gamma\vdash\letbe{\pair{X,x^{\tau_1}}}{M}{N}:\tau_2}
      {\Gamma\vdash M:\exists X.\tau_1 &
       \Gamma,x:\tau_1\vdash N:\tau_2 & X\!\not\in\! FTV(\Gamma,\tau_2)}
$$
Again, we employ the standard $\beta\eta$-axioms.
$$
\begin{array}{rcll}
(\lambda x^\sigma.M)\,N &=& M[N/x] \\
\lambda x^\sigma.M\,x &=& M ~~(x\not\in FV(M)) \\
\letbe{\pair{x,y}}{\pair{L,M}}{N} &=& N[L/x,M/y] \\
\letbe{\pair{x,y}}{M}{N[\pair{x,y}/z]} &=& N[M/z] \\
\letbe{\pair{X,x}}{\pair{\tau,M}}{N} &=& N[\tau/X,M/x] \\
\letbe{\pair{X,x}}{M}{N[\pair{X,x}/z]} &=& N[M/z] 
\end{array}
$$

\section{CPS translation}

\subsection{The CPS translation}

We present a call-by-name CPS translation which can be considered as an 
extension of that introduced by Streicher \cite{HS02,SR98,Sel01}
(rather than the translations by Plotkin \cite{Plo75}, Parigot \cite{Par97}
or Fujita \cite{Fuj04} which introduce extra negations and 
do not respect extensionality).
$$
\begin{array}{rcl}
X^\circ &=& X\\
(\sigma_1\rightarrow\sigma_2)^\circ &=& \neg\sigma_1^\circ\wedge\sigma_2^\circ\\
(\forall X.\sigma)^\circ &=& \exists X.\sigma^\circ\\
\\
\cps{x^\sigma} &=& x^{\neg\sigma^\circ}\\
\cps{\lambda x^{\sigma_1}.M^{\sigma_2}} &=& 
 \lambda\pair{x^{\neg\sigma_1^\circ},k^{\sigma_2^\circ}}.\cps{M}\,k\\
\cps{M^{\sigma_1\rightarrow\sigma_2}\,N^{\sigma_1}} &=& 
 \lambda k^{\sigma_2^\circ}.\cps{M}\,\pair{\cps{N},k}\\
\cps{\Lambda X.M^\sigma} &=& 
 \lambda\pair{X,k^{\sigma^\circ}}.\cps{M}\,k\\
\cps{M^{\forall X.\sigma_1}\,\sigma_2} &=& 
 \lambda k^{\sigma_1[\sigma_2/X]^\circ}.\cps{M}\,\pair{\sigma_2^\circ,k}\\
\cps{\mu\alpha^{\sigma_1}.\name{\beta^{\sigma_2}}M^{\sigma_2}} &=&
 \lambda\alpha^{\sigma_1^\circ}.\cps{M}\,\beta
\end{array}
$$
where
$$
\begin{array}{lcl}
\lambda\pair{x^\sigma,y^\tau}.M &\equiv& 
 \lambda z^{\sigma\wedge\tau}.\letbe{\pair{x^\sigma,y^\tau}}{z}{M}\\
\lambda\pair{X,y^\tau}.M &\equiv& 
 \lambda z^{\exists X.\tau}.\letbe{\pair{X,y^{\tau}}}{z}{M}\\
\end{array}
$$

\subsection{Soundness}

The type soundness follows from a straightforward induction.
\begin{proposition}[type soundness]
$$\Gamma\vdash M:\sigma~|~\Delta~\Longrightarrow~\neg\Gamma^\circ,\Delta^\circ\vdash \cps{M}:\neg\sigma^\circ$$
where $\neg\Gamma^\circ$ is 
$x_1:\neg\sigma_1^\circ,\dots,x_m:\neg\sigma_m^\circ$ 
when $\Gamma$ is $x_1:\sigma_1,\dots,x_m:\sigma_m$, and 
$\Delta^\circ=\alpha_1:\sigma_1^\circ,\dots,\alpha_n:\sigma_n^\circ$
for $\Delta=\alpha_1:\sigma_1,\dots,\alpha_n:\sigma_n$.
\qed
\end{proposition}
Note that
$(\sigma[\tau/X])^\circ\equiv\sigma^\circ[\tau^\circ/X]$,
$\cps{M[N/x]}\equiv\cps{M}[\cps{N}/x]$, and also
$\cps{M[\sigma/X]}\equiv\cps{M}[\sigma^\circ/X]$ hold.
Then we have the equational soundness:
\begin{proposition}[equational soundness]
$$\Gamma\vdash M=N:\sigma~|\Delta~\Longrightarrow~\neg\Gamma^\circ,\Delta^\circ\vdash \cps{M}=\cps{N}:\neg\sigma^\circ$$
\qed
\end{proposition}
In addition, we have the definability result:
\begin{proposition}[fullness]
$$\hbox to 79pt{\hfil}\neg\Gamma^\circ,\Delta^\circ\vdash N:\neg\sigma^\circ
\Longrightarrow N=\cps{M}~\mathrm{for~some}~\Gamma\vdash
M:\sigma~|~\Delta\hbox to 79pt{\hfil}\qEd$$
\end{proposition}
This can be proved by providing an inverse translation of the
CPS translation, 
so that
$$
\begin{array}{rcl}
\neg\Gamma^\circ,\Delta^\circ\vdash P:\neg\sigma^\circ
&\Longrightarrow&
\Gamma\vdash P^{-1}:\sigma~|~\Delta\\
\neg\Gamma^\circ,\Delta^\circ\vdash C:\sigma^\circ
&\Longrightarrow&
\Gamma\vdash C^{-1}[-^\sigma]:\bot~|~\Delta\\
\neg\Gamma^\circ,\Delta^\circ\vdash A:R
&\Longrightarrow&
\Gamma\vdash A^{-1}:\bot~|~\Delta
\end{array}
$$
hold, where
$$
\begin{array}{llll}
\mathrm{Program}:\neg\sigma^\circ & P &\!::=\!& x~|~\lambda k.A\\
\mathrm{Continuation}:\sigma^\circ\!\! & C &\!::=\!& k~|~\pair{P,C}~|~\pair{\sigma^\circ,C}~|~\letbe{\pair{x,k}}{C}{C}~|~\letbe{\pair{X,k}}{C}{C}\\
\mathrm{Answer}:R & A &\!::=\!& P\,C~|~\letbe{\pair{x,k}}{C}{A}~|~
\letbe{\pair{X,k}}{C}{A}\\
\end{array}
$$
as follows.
$$
\begin{array}{rcl}
x^{-1} &=& x\\
(\lambda k^{\sigma^\circ}.A)^{-1} &=& \mymu k^\sigma.A^{-1}\\
\\
k^{-1} &=& \myname{k}[-]\\
\pair{P,C}^{-1} &=& C^{-1}[-\,P^{-1}]\\
\pair{\sigma^\circ,C}^{-1} &=& C^{-1}[-\,\sigma]\\
(\letbe{\pair{x,k}}{C_1}{C_2})^{-1} &=& C_1^{-1}[\lambda x.\mu k.C_2^{-1}[-]]\\
(\letbe{\pair{X,k}}{C_1}{C_2})^{-1} &=& C_1^{-1}[\Lambda X.\mu k.C_2^{-1}[-]]\\
\\
(P\,C)^{-1} &=& C^{-1}[P^{-1}]\\
(\letbe{\pair{x,k}}{C}{A})^{-1} &=& C^{-1}[\lambda x.\mymu k.A^{-1}]\\
(\letbe{\pair{X,k}}{C}{A})^{-1} &=& C^{-1}[\Lambda X.\mymu k.A^{-1}]
\\
\end{array}
$$
This can be considered as
a ``continuation-grabbing style transformation'' in the sense of Sabry
\cite{Sab96}.
It follows that for any 
$\neg\Gamma^\circ,\Delta^\circ\vdash M:\neg\sigma^\circ$
there exists 
$\neg\Gamma^\circ,\Delta^\circ\vdash P:\neg\sigma^\circ$
generated by this grammar such that $P= M$
--- it suffices to take the $\beta$-normal form \cite{Fuj04}. 
Moreover we can routinely show that $\cps{P^{-1}}=P$.
Thus the CPS translation enjoys fullness: 
all terms are definable modulo the provable equality.
This definability is important 
for relating the parametricity principles for the source
and target calculi.

\subsection{A semantic explanation}

Here is a short explanation of why this CPS translation works,
intended for readers with category theoretic background ---
on the ``categories of continuations'' construction \cite{HS02,SR98,Sel01},
and on fibrations for polymorphic type theories \cite{Jac99}. 
As a response category $\calC$ with a response object $R$
induces a control category $R^\calC$ with $R^\calC(X,Y)=\calC(R^X,R^Y)$,
a fibred response category with finite products and simple coproducts
(for existential quantifiers) 
induces a fibred control category with finite products and simple
products (for universal quantifiers).
Let us write $\calC_\Gamma$ for the response category over the
type-context $\Gamma$. We assume that the weakening functor
$\pi^*:\calC_\Gamma\rightarrow\calC_{\Gamma\times A}$ has a 
left adjoint 
$\exists_A:\calC_{\Gamma\times A}\rightarrow\calC_\Gamma$
subject to the Beck-Chevalley condition.
Thus
$$
\calC_{\Gamma\times A}(X,\pi^*(Y))\simeq\calC_\Gamma(\exists_A(X),Y)
$$
We then have
$$
\begin{array}{rcl}
R^{\calC_{\Gamma\times A}}(\pi^*(X),Y)
&=&
\calC_{\Gamma\times A}(R^{\pi^*(X)},R^Y)\\
&\simeq&
\calC_{\Gamma\times A}(Y,R^{R^{\pi^*(X)}})\\
&\simeq&
\calC_{\Gamma\times A}(Y,\pi^*(R^{R^X}))\\
&\simeq&
\calC_\Gamma(\exists_A(Y),R^{R^X})\\
&\simeq&
\calC_\Gamma(R^X,R^{\exists_A(Y)})\\
&=&
R^{\calC_\Gamma}(X,\exists_A(Y))
\end{array}
$$
Hence $\pi^*$, regarded as the weakening functor from
$R^{\calC_\Gamma}$ to $R^{\calC_{\Gamma\times A}}$,
has a right adjoint given by $\exists_A$, which can be used 
for interpreting the universal quantifier. 
Our CPS transformation is essentially a syntactic 
interpretation of this semantic construction.

\section{CPS semantics with the parametric target calculus}

\subsection{Parametricity for the target calculus}

As the target calculus can be seen as a subset of $\lambda2$
(via the standard encoding of the conjunctions and
existential types), we can define
the relational parametricity for the target calculus in the same way
as for $\lambda2$, e.g. logic 
for parametricity \cite{PA93,Tak98}, system R \cite{ACC93}, or 
system P \cite{Dun02}.
One may directly define the parametricity principle 
(often called the simulation principle) for the existential type, 
see for example \cite{PA93}.

In this paper we only consider the relations constructed 
from the graphs of terms-in-context, identity, and $\sigma^*$'s 
obtained by the following construction, which we shall call 
``admissible relations''.

Among admissible relations, the most fundamental are the graph relations.
Given a term $f\,(x):\tau_2$ with a free variable $x:\tau_1$
we define its graph relation 
$\pair{x\vdash f\,(x)}:\tau_1\leftrightarrow\tau_2$
($\pair{f}$ for short) by
$u\,\pair{f}\,v$ iff $f\,(u)=v$. 

Given a type $\tau$ whose free type variables are included in
$X_1,\dots,X_n$ and admissible relations
$s_1:\tau_1\leftrightarrow\tau_1',\dots,
 s_n:\tau_n\leftrightarrow\tau_n'$,
we define an admissible relation $\tau^*$ as follows.
\begin{itemize}
\item $X_i^*=s_i:\tau_i\leftrightarrow\tau_i'$
\item $R^*$ is the identity relation on the terms of type $R$
\item $(\neg\tau)^*:\neg\tau[\tau_1/X_1,\dots]\leftrightarrow\neg\tau[\tau_1'/X_1,\dots]$ is the relation so that
      $f\,(\neg\tau)^*\,g$ iff $x\,\tau^*\,y$ implies $f\,x\,R^*\,g\,y$
      (hence $f\,x=g\,y$)
\item $(\tau\wedge\tau')^*:(\tau\wedge\tau')[\tau_1/X_1,\dots]\leftrightarrow
  (\tau\wedge\tau')[\tau_1'/X_1,\dots]$ is the relation so that
      $u\,(\tau\wedge\tau')^*\,v$ iff
      $u=\pair{x,x'}$, $v=\pair{y,y'}$ and $x\,\tau^*\,y$, $x'\,\tau'^*\,y'$
\item $(\exists X.\tau)^*:\exists X.\tau[\tau_1/X_1,\dots]\leftrightarrow\exists X.\tau[\tau_1'/X_1,\dots]$ is the relation so that
      $u\,(\exists X.\tau)^*\,v$ iff
      $u=\pair{\tau',x}$, $v=\pair{\tau'',y}$ and
      $x\,\tau[r/X]^*\,y$ for some admissible $r:\tau'\leftrightarrow\tau''$
\end{itemize}
In the last case, the relation $\tau[r/X]^*:\tau[\tau'/X]\leftrightarrow\tau[\tau''/X]$ is defined as $\tau^*$ with $X^*=r$.
One may further define admissible relations
$\neg r$, $r\wedge s$ and $\exists X.r$ for admissible $r$, $s$, so that
$(\neg\tau)^*=\neg\tau^*$, $(\tau\wedge\tau')^*=\tau^*\wedge\tau'^*$
and $(\exists X.\tau)^*=\exists X.\tau^*$ hold.

Let $id_\tau:\tau\leftrightarrow\tau$ 
be the identity relation on the terms of type $\tau$.
The relational parametricity asserts that, for 
any $\tau$ whose free type variables are included in
$X_1,\dots,X_n$ and $\tau_1,\dots,\tau_n$, 
$M:\tau[\tau_1/X_1,\dots,\tau_n/X_n]$ implies 
$M\,\tau^*\,M$ with $s_i=id_{\tau_i}$.

Its consistency follows immediately from 
that of the parametricity for $\lambda2$.

\begin{proposition}
As consequences of the parametricity,
we can derive:
\begin{enumerate}
\item $\exists X.X$ gives a terminal object $\top$ with a unique inhabitant
  $*$, so that for any $M:\top$ we have $M=*$.
\item $\exists X.\neg(\tau\wedge X)\wedge X$
(which could be rewritten as $\exists X.(X\rightarrow\neg\tau)\wedge X$)
gives a final coalgebra $\nu X.\neg\tau$ of $\Lambda X.\neg\tau$ where
$X$ only occurs negatively in $\tau$.
\item (as an instance of the last case) the isomorphism
$\exists X.\neg(\tau\wedge X)\wedge X\simeq\neg\tau$ holds 
if $X$ does not occur freely in $\tau$.\qed
\end{enumerate}

\end{proposition}
Their proofs are standard, cf. papers cited above 
\cite{PA93,ACC93,Tak98,Dun02}. 

Below we will see the implications of these parametricity results 
on the target calculus. We refer to the $\lambda\mu2$-theory induced
by the CPS translation into this parametric target calculus as
$\lambda\mu2^P$.

\subsection{The falsity type}

As a first example, let us consider the 
falsity type $\bot=\forall X.X$ in $\lambda\mu2$.
We have
$$
\bot^\circ=(\forall X.X)^\circ=\exists X.X\simeq\top
$$
and 
$$
(\sigma\rightarrow\bot)^\circ=
\neg(\sigma^\circ\wedge\bot^\circ)\simeq
\neg(\sigma^\circ\wedge\top)\simeq
\neg\sigma^\circ
$$
Since $\exists X.X$ is terminal (with a unique inhabitant $*$)
in the parametric target calculus, we obtain 
$\cps{\mymu\alpha^\sigma.M}=\lambda\alpha^{\sigma^\circ}.\cps{M}*$ and
$\cps{\myname{\beta}M}=\lambda u^{\exists X.X}.\cps{M}\beta$,
which coincide with Streicher's translation. As a consequence, 
the following equations on the named terms and $\mymu$-abstractions 
are all validated in $\lambda\mu2^P$.
$$
\begin{array}{rcll}
(\mymu\alpha^{\sigma_1\rightarrow\sigma_2}.M)\,N &=&
\mymu\beta^{\sigma_2}.M[\name{\beta}(-\,N)/\name{\alpha}(-)] & \\
(\mymu\alpha^{\forall X.\sigma_1}.M)\,\sigma_2 &=&
\mymu\beta^{\sigma_1[\sigma_2/X]}.M[\name{\beta}(-\,\sigma_2)/\name{\alpha}(-)] & \\
\myname{\alpha'}(\mymu\alpha^\sigma.M) &=& M[\alpha'/\alpha] & \\
\myname{\alpha^\bot}M &=& M & 
\end{array}
$$
Thus the type $\bot$ serves as the falsity type
as found in some formulation of the $\lambda\mu$-calculus.
In addition, we can show that 
$(\sigma,\,C_\sigma:((\sigma\rightarrow\bot)\rightarrow\bot)\rightarrow\sigma)$ is an algebra of the double-negation monad
$((-)\rightarrow\bot)\rightarrow\bot$ on the term model.

\subsection{Initial algebra?}

A more substantial example is the ``initial algebra''
$\mu X.F[X]=\forall X.(F[X]\rightarrow X)\rightarrow X$,
with $X$ positive in $F[X]$
(here we see an unfortunate clash of $\mu$'s for the name-binding and for
the fixed-point on types, but this should not cause any serious
problem). We calculate:
$$
\begin{array}{rcl}
(\mu X.F[X])^\circ
&=&
(\forall X.(F[X]\rightarrow X)\rightarrow X)^\circ\\
&=&
\exists X.\neg(\neg F[X]^\circ\wedge X)\wedge X\\
&\simeq&
\nu X.\neg\neg F[X]^\circ\\
&\simeq&
\neg\neg F[X]^\circ[\nu X.\neg\neg F[X]^\circ/X]\\
&\simeq&
\neg\neg F[X]^\circ[(\mu X.F[X])^\circ/X]\\
&=&
\neg\neg(F[\mu X.F[X]])^\circ\\
&\simeq&
((F[\mu X.F[X]]\rightarrow\bot)\rightarrow\bot)^\circ
\end{array}
$$
This suggests that $\mu X.F[X]$ is isomorphic not to
$F[\mu X.F[X]]$ but to its double negation
$(F[\mu X.F[X]]\rightarrow\bot)\rightarrow\bot$. 
One might think that this contradicts the standard 
experience on $\lambda2$
with parametricity, where we have an isomorphism 
$\mathsf{in}:F[\mu X.F[X]]\rightarrow\mu X.F[X]$.
Since $\lambda\mu2$ subsumes $\lambda2$, we have 
this $\mathsf{in}$ in $\lambda\mu2$ too; however, it should not
be an isomorphism, regarding the CPS interpretation above
(otherwise it causes a degeneracy).
The truth is that, in $\lambda\mu2^P$,
the term 
$$
\mathsf{in}^\sharp=\lambda m.\mymu\alpha.m\,(\lambda x.\myname{\alpha}(\mathsf{in}\,x))
~:~((F[\mu X.F[X]]\rightarrow\bot)\rightarrow\bot)\rightarrow
 \mu X.F[X]
$$
is an isomorphism.
It still is not an initial algebra of $(F[-]\rightarrow\bot)\rightarrow\bot$; 
we shall further consider this issue later.
For now, we shall emphasize that the parametricity principle
for $\lambda2$ should not be used for $\lambda\mu2$, at least 
without certain constraint --- otherwise $\mathsf{in}$ would be an
isomorphism, hence a degeneracy follows (because we have
$(\sigma\rightarrow\bot)\rightarrow\bot\simeq\sigma$ for every $\sigma$).

\subsection{Other impredicative encodings}

Recall other impredicative encodings of logical connectives:
$$
\begin{array}{rcl}
\top &=& \forall X.X\rightarrow X\\
\sigma_1\wedge\sigma_2 &=& \forall X.(\sigma_1\rightarrow\sigma_2\rightarrow X)\rightarrow X\\
\sigma_1\vee\sigma_2 &=&
\forall X.(\sigma_1\rightarrow X)\rightarrow(\sigma_2\rightarrow X)\rightarrow X\\
\exists X.\sigma &=& \forall Y.(\forall X.(\sigma\rightarrow Y))\rightarrow Y\\
\end{array}
$$
Their CPS translations into the parametric target calculus satisfy:
$$
\begin{array}{rcl}
\top^\circ
&\simeq&
R\\
(\sigma_1\wedge\sigma_2)^\circ
&\simeq&
\neg(\neg{\sigma_1}^\circ\wedge\neg\sigma_2^\circ)\\
(\sigma_1\vee\sigma_2)^\circ
&\simeq&
\neg\neg\sigma_1^\circ\wedge\neg\neg\sigma_2^\circ\\
(\exists X.\sigma)^\circ
&\simeq&
\neg\exists X.\neg\sigma^\circ\\
\end{array}
$$
As easily seen, these defined logical connectives in the source 
calculus do not obey 
the standard universal properties as in the parametric models of $\lambda2$. 
In short, they are all ``double-negated'', hence amount to some
classical encodings:
\begin{itemize}
\item $\sigma_1\wedge\sigma_2$ is not a cartesian product 
of $\sigma_1$ and $\sigma_2$, but isomorphic to
$(\sigma_1\rightarrow\sigma_2\rightarrow\bot)\rightarrow\bot$.
It is possible to add cartesian product types
$\sigma_1\times\sigma_2$ to $\lambda\mu2$, but then 
we also need to add coproduct types $\tau_1+\tau_2$ to the target calculus,
so that $(\sigma_1\times\sigma_2)^\circ=\sigma_1^\circ+\sigma_2^\circ$ and
$\sigma_1\wedge\sigma_2\simeq\neg\neg(\sigma_1\times\sigma_2)$.
\item $\top$ is not a terminal object, but isomorphic to
$\bot\rightarrow\bot$. 
We can add a terminal object $1$ to $\lambda\mu2$ and an initial 
object $0$ to the target, so that $1^\circ=0$ and $\top\simeq \neg\neg 1$.
\item $\sigma_1\vee\sigma_2$ is not a coproduct of $\sigma_1$ and $\sigma_2$,
but isomorphic to $(\sigma_1\rightarrow\bot)\rightarrow(\sigma_2\rightarrow\bot)\rightarrow\bot$.
If there is a coproduct $\sigma_1+\sigma_2$, then it should follow
that $\sigma\vee\tau\simeq\neg\neg(\sigma+\tau)$.
On the other hand, it is not possible to enrich $\lambda\mu2$
with an initial object without a degeneracy, 
cf. Selinger's note on control categories \cite{Sel03}.
Alternatively we might add the ``classical disjunction types'' 
$\sigma_1\wp\sigma_2$ \cite{Sel01}
with $(\sigma_1\wp\sigma_2)^\circ=\sigma_1^\circ\wedge\sigma_2^\circ$ --- hence 
$\sigma_1\rightarrow\sigma_2\simeq\neg\sigma_1\wp\sigma_2$ and
$\sigma_1\vee\sigma_2\simeq\neg\neg\sigma_1\wp\neg\neg\sigma_2$.
We note that $\bot=\forall X.X$ serves as the unit of this classical
disjunction.
\item $\exists X.\sigma$ does not work as the existential type; 
it is isomorphic to $\neg\forall X.\neg\sigma$.
\end{itemize}

\subsection{Answer-type polymorphism}

Note that the answer type $R$ has been considered just as a constant 
with no specific property. In fact we could have used any type for $R$ ---
Everything is defined polymorphically regarding $R$.
Thus we can apply the ``answer-type polymorphism'' principle 
(cf. \cite{Thi04}):
in particular, a closed term of type $\sigma$ in $\lambda\mu2$
can be considered to be sent to a $\lambda2$-term of type 
$\forall R.\neg\sigma^\circ$.
This way of reasoning goes behind the parametricity principle for 
our target calculus, but it is justified by the parametricity of
$\lambda2$. 

For instance, consider the type 
$\top=\forall X.X\rightarrow X$ of $\lambda\mu2$. 
We have 
$$
\begin{array}{rcl}
\forall R.\neg\top^\circ
&\simeq&
\forall R.\neg R\\
&\simeq&
\forall R.R\rightarrow R\\
&\simeq&
1
\end{array}
$$
in $\lambda2$ with parametricity. 
This means that,
although $\top$ is not a terminal object
in $\lambda\mu2$, it has a unique closed inhabitant.
Similarly, we have $\forall R.\neg\bot^\circ\simeq\forall R.R\simeq 0$,
thus we see that there is no closed inhabitant of $\bot$ in $\lambda\mu2$.

However, such reasonings based on the answer-type
polymorphism become much harder for more complicated types.
The force of answer-type polymorphism in this setting seems
still not very obvious.

\section{Focal parametricity}

We have seen that the CPS semantics with respect to the 
target calculus with relational parametricity induces a
reasonable equational theory $\lambda\mu2^P$. 
However, here the parametricity is used rather indirectly, via the
CPS translation.
We now consider a notion of 
parametricity which is directly available within $\lambda\mu2$.

\subsection{CPS translating relations}

The key of formulating the relational parametricity is 
the use of graph relations of terms (considered as representing a functional
relation): without graph relations,
relational parametricity reduces to just the basic lemma 
of the (second-order) logical relations. On the other hand,
it does not have to allow all terms to be used for constructing 
relations. In fact, in linear parametricity \cite{Plo93}
only linear (or strict) maps are allowed to be used for constructing 
graph relations, and this choice allows a weaker notion of 
parametricity which can accommodate recursion. 
Naturally, we are led to look for a characterisation of $\lambda\mu$-terms 
which can be used for graph relations without breaking the soundness 
with respect to the CPS semantics into the parametric target calculus. 

Now suppose that we are allowed to use the graph relation
$\pair{f}:\sigma_1\leftrightarrow\sigma_2$ of a term 
$x:\sigma_1\vdash f(x):\sigma_2$. To ensure the soundness of the
use of this graph relation, we shall consider the 
CPS translation of such relations. For instance, we 
hope that $\pair{f}$ will be sent to a relation between 
types $\sigma_2^\circ$ and $\sigma_1^\circ$ in the target 
calculus. However, since 
$x:\neg\sigma_1^\circ\vdash\cps{f(x)}:\neg\sigma_2^\circ$,
we have some relation $\sigma_2^\circ\leftrightarrow\sigma_1^\circ$
only when $\cps{f(x)}=\lambda k.x\,(g\,(k))$
for some $k:\sigma_2^\circ\vdash g(k):\sigma_1^\circ$ in the target
calculus. If there is such $g$, we can complete the translation 
of the relations and reduce the parametricity principle on $\lambda\mu2$
to the parametricity on the target calculus.

Fortunately, there is a way to characterise such ``translatable''
$f$'s in the $\lambda\mu$-calculus without performing the 
CPS-translation (modulo a technical assumption on the CPS-target,
known as ``equalising requirement'' \cite{Mog91}). 
It is the notion of ``focus'', which we now recall below.

\subsection{Focus}

\begin{definition}
A $\lambda\mu2$-term $M:\sigma_1\rightarrow\sigma_2$ is called
{\em focal} if it is an algebra morphism from $(\sigma_1,C_{\sigma_1})$
to $(\sigma_2,C_{\sigma_2})$, i.e. the following diagram commutes.
$$
\begin{diagram}
\node{(\sigma_1\rightarrow\bot)\rightarrow\bot}
 \arrow{s,l}{C_{\sigma_1}}
 \arrow[2]{e,t}{(M\rightarrow\bot)\rightarrow\bot}
\node{}
\node[2]{(\sigma_2\rightarrow\bot)\rightarrow\bot}
 \arrow{s,r}{C_{\sigma_2}}\\
\node{\sigma_1}
 \arrow[3]{e,b}{M}
\node{}
\node[2]{\sigma_2}
\end{diagram}
$$
That is:
$$
 M\,(\mymu\alpha^{\sigma_1}.k\,(\lambda x^{\sigma_1}.\myname{\alpha}x))
 =
 \mymu\beta^{\sigma_2}.k\,(\lambda x^{\sigma_1}.\myname{\beta}(M\,x))
 :\sigma_2
$$
holds for any $k:(\sigma_1\rightarrow\bot)\rightarrow\bot$.%
\footnote{
In \cite{HS02}, a focal map from $\sigma_1$ to $\bot$ is called
a ``C-term of type $\sigma_1$''. C-terms of type $\sigma_1$ with a free 
name of $\sigma_2$ correspond to focal maps from $\sigma_1$ to $\sigma_2$,
thus these notions (and the associated constructions of
the CPS target categories via C-terms (C-maps) \cite{HS02} and via
focus \cite{Sel01})
are essentially the same.}
\end{definition}
In any $\lambda\mu2$-theory, 
focal terms compose, and the identity $\lambda x^\sigma.x$ is
obviously focal. So, the (equivalence classes of) focal maps
form a category. Hereafter we shall call it the {\em focus}
of the $\lambda\mu2$-theory.

While this characterisation of focal maps is concise and 
closely follows the semantic considerations in \cite{Sel01,KH03},
there is a subtle problem; the $\beta\eta$-axioms of 
$\lambda\mu2$ are too weak to establish the focality of
some important terms. 
This is because 
we have used the polymorphic feature of $\lambda\mu2$ for expressing 
$C_\sigma$ --- it involves the falsity type $\bot=\forall X.X$,
but the axioms of $\lambda\mu2$ do not guarantee that $\bot$ does
work properly. If there were not sufficiently many focal maps,
the parametricity principle restricted on focal maps
would be useless.

To see this issue more
clearly, we shall look at another ``classical'' combinator (the Peirce law)
$$
P_{\sigma_1,\sigma_2}=
\lambda m.
 \mu\alpha^{\sigma_1}.\name{\alpha}(m\,(\lambda x^{\sigma_1}.\mu\beta^{\sigma_2}.\name{\alpha}x))
~:~((\sigma_1\rightarrow\sigma_2)\rightarrow\sigma_1)\rightarrow\sigma_1
$$
which does not make use of polymorphism, and
the ``abort'' map (Ex Falso Quodlibet)
$$
A_{\sigma}=\lambda x^\bot.x\,\sigma:\bot\rightarrow\sigma
$$
which is defined without the classical feature.
It is well known that the double-negation elimination 
is as expressible as the Peirce law together with Ex Falso Quodlibet,
see e.g. \cite{AH03}. This is also the case at the level of
(uniformity of) proofs.
Let us say that $M:\sigma_1\rightarrow\sigma_2$ is {\em repeatable}
if 
$$
\begin{diagram}
\node{(\sigma_1\rightarrow\sigma_3)\rightarrow\sigma_1}
 \arrow{s,l}{P_{\sigma_1,\sigma_3}}
 \arrow[2]{e,t}{(M\rightarrow\sigma_3)\rightarrow M}
\node{}
\node[2]{(\sigma_2\rightarrow\sigma_3)\rightarrow\sigma_2}
 \arrow{s,r}{P_{\sigma_2,\sigma_3}}\\
\node{\sigma_1}
 \arrow[3]{e,b}{M}
\node{}
\node[2]{\sigma_2}
\end{diagram}
$$
commutes for each $\sigma_3$; and {\em discardable} if
$$
\begin{diagram}
\node[2]{\bot}
 \arrow{sw,t}{A_{\sigma_1}}
 \arrow{se,t}{A_{\sigma_2}}\\
\node{\sigma_1}
 \arrow[2]{e,b}{M}
\node[2]{\sigma_2}
\end{diagram}
$$
commutes.
\begin{proposition}
In a $\lambda\mu2$-theory, 
$M:\sigma_1\rightarrow\sigma_2$ is focal if and only if it is
both repeatable and discardable.
\qed
\end{proposition}
We note that the corresponding result in the call-by-value
setting has been observed by F\"uhrmann 
\cite{Fuh02} as the characterisation of algebraic values
as repeatable discardable expressions. Here we follow his terminology.

This reformulation allows us to see that 
only the second diagram of $A$'s involves the polymorphically 
defined $\bot$ and needs to be justified by additional conditions.%
\footnote{This problem was overlooked in the preliminary version of
this paper \cite{Has05} where it was wrongly assumed
that repeatability alone would imply focality.}
On the other hand, the first diagram of $P$'s is not problematic,
as it does not make use of polymorphism at all.

\subsection{Additional axioms}

To this end, we add more axioms to $\lambda\mu2$ before 
thinking about parametricity. They are
\begin{enumerate}
\item $\lambda x^{\sigma_1\rightarrow\sigma_2}.x\,N:
  (\sigma_1\rightarrow\sigma_2)\rightarrow\sigma_2$ is discardable 
  for any $N:\sigma_1$
\item $\lambda x^{\forall X.\sigma}.x\,\sigma_1:
  \forall X.\sigma\rightarrow\sigma[\sigma_1/X]$ is discardable
  for any $\sigma$ and $\sigma_1$
\item $\lambda x^{\sigma}.\myname{\alpha}x:\sigma\rightarrow\bot$ is discardable
  for any $\alpha:\sigma$
\end{enumerate}
which are equivalent to asking
\begin{enumerate}
\item $M\,(\sigma_1\rightarrow\sigma_2)\,N=M\,\sigma_2$ for any
  $M:\bot$ and $N:\sigma_1$
\item $M\,(\forall X.\sigma)\,\sigma_1=M\,(\sigma[\sigma_1/X])$
  for $M:\bot$
\item $\myname{\alpha}(M\,\sigma)=M$ for $M:\bot$ and $\alpha:\sigma$
  ($\alpha\not\in FN(M)$)
\end{enumerate}
and also equivalent to 
\begin{enumerate}
\item
$(\mymu\alpha^{\sigma_1\rightarrow\sigma_2}.M)\,N =
\mymu\beta^{\sigma_2}.M[\name{\beta}(-\,N)/\name{\alpha}(-)]$
\item
$(\mymu\alpha^{\forall X.\sigma_1}.M)\,\sigma_2 =
\mymu\beta^{\sigma_1[\sigma_2/X]}.M[\name{\beta}(-\,\sigma_2)/\name{\alpha}(-)]$
\item
$\myname{\alpha'}(\mymu\alpha^\sigma.M) = M[\alpha'/\alpha]$
\end{enumerate}
Note that $\lambda\mu2^P$ discussed in Section 4 satisfies these 
conditions.
Also we shall note that $\lambda x^{\sigma_1\rightarrow\sigma_2}.x\,N$, 
$\lambda x^{\forall X.\sigma}.x\,\sigma_1$, 
$\lambda x^{\bot}.\myname{\alpha}(x\,\sigma)$ 
are all repeatable in $\lambda\mu2$. Together with these additional
axioms, they become focal. 
(Alternatively, we could have $\bot$ as a type constant and assume 
the standard axiomatization of $\lambda\mu$-calculus with the falsity type
\cite{HS02,Sel01}
--- in that case
$C$ is defined without polymorphism, and 
this problem disappears.)


Below we develop the focal parametricity principle 
on top of $\lambda\mu2$ with these additional axioms.

\subsection{A parametricity principle for
\texorpdfstring{\boldmath$\lambda\mu2$}{}}

Given a focal $f:\sigma_1\rightarrow\sigma_2$
we define its graph relation 
$\pair{f}:\sigma_1\leftrightarrow\sigma_2$
by
$u\,\pair{f}\,v$ iff $f\,u=v$.  
Also, let $id_\sigma:\sigma\leftrightarrow\sigma$ 
be the identity relation on the terms of type $\sigma$.
In this paper we only consider the relations given by
the graphs of focal maps, identity, and $\sigma^*$'s obtained
by the following construction, which we shall call 
``focal relations''.

Given a type $\sigma$ whose free type variables are included in
$X_1,\dots,X_n$ and focal relations
$s_1:\sigma_1\leftrightarrow\sigma_1',\dots,
 s_n:\sigma_n\leftrightarrow\sigma_n'$,
we define a focal relation $\sigma^*$ as follows.
\begin{itemize}
\item $X_i^*=s_i:\sigma_i\leftrightarrow\sigma_i'$
\item $(\sigma\!\rightarrow\!\sigma')^*\!:\!(\sigma\!\rightarrow\!\sigma')[\sigma_1/X_1,\dots]\!\leftrightarrow\!(\sigma\!\rightarrow\!\sigma')[\sigma_1'/X_1,\dots]$ is the relation so that
      $f\,(\sigma\rightarrow\sigma')^*\,g$ iff 
      $x\,\sigma^*\,y$ implies $(f\,x)\,\sigma'^*\,(g\,y)$
\item $(\forall X.\sigma)^*:\forall X.\sigma[\sigma_1/X_1,\dots]\leftrightarrow\forall X.\sigma[\sigma_1'/X_1,\dots]$ is the relation so that
      $u\,(\forall X.\sigma)^*\,v$ iff
      $(u\,\sigma')\,\sigma[r/X]^*\,(v\,\sigma'')$ holds
      for any focal relation $r:\sigma'\leftrightarrow\sigma''$
\end{itemize}

The focal relational parametricity asserts that, for 
any $\sigma$ whose free type variables are included in
$X_1,\dots,X_n$, 
$M:\sigma[\sigma_1/X_1,\dots,\sigma_n/X_n]$ implies 
$M\,\sigma^*\,M$ with $s_i=id_{\sigma_i}$.

Thus the only departure from the standard parametricity principle
is the condition that the graph relation construction is allowed 
only on focal maps.
Note that this restriction is necessary;
if we apply parametricity to
polymorphic terms $\Lambda X.C_X$ or $\Lambda X.P_{X,\sigma}$,
we will get the naturality diagrams above for any term
which is allowed to be used for the graph relation construction.

\subsection{On consistency and soundness}

The consistency of focal parametricity 
(in the sense that the equational theory of $\lambda\mu2$ with 
focal parametricity is not trivial)
follows from the fact
that there are non-trivial parametric models of $\lambda2$ in which there is
an object $R$ so that the continuation monad $T\tau=R^{R^\tau}$ satisfies
the ``equalising requirement'' \cite{Mog91}, i.e. each component
$\eta_\tau:\tau\rightarrow T\tau$ of its unit is an equaliser of 
$\eta_{T\tau}$ and $T\eta_\tau$. 
(Here we employ the syntax of the CPS target calculus as 
an internal language for such models, where the CPS translation
is considered to give a semantic interpretation.)
In such models, for any focal term $f:\sigma_1\rightarrow\sigma_2$,
there exists a unique
$y:\sigma_2^\circ\vdash g(y):\sigma_1^\circ$
such that $\cps{f\,x}=\lambda y.x\,(g(y))$ (cf. \cite{Sel01}). 

Using this fact, 
given a focal relation $r:\sigma_1\leftrightarrow\sigma_2$,
we construct an admissible relation 
$r^\circ:\sigma_2^\circ\leftrightarrow\sigma_1^\circ$ as follows.
For a graph relation $\pair{f}:\sigma_1\leftrightarrow\sigma_2$,
we let 
$\pair{f}^\circ=\pair{g}:\sigma_2^\circ\leftrightarrow\sigma_1^\circ$
where $g$ is the unique map as given above.
For $\sigma^*$, $\sigma^{*\circ}$ is defined by straightforward induction:
$(\sigma\rightarrow\sigma')^{*\circ}=
 \neg\sigma^{*\circ}\wedge\sigma'^{*\circ}$,
$(\forall X.\sigma)^{*\circ}=\exists X.\sigma^{*\circ}$
(where the parameter relations $s_i$ are replaced by $s_i^\circ$).

\begin{theorem}
In such a model, given a focal relation $r:\sigma_1\leftrightarrow\sigma_2$,
$M\,r\,N$ implies $\cps{N}\,\neg r^\circ\,\cps{M}$.
\qed
\end{theorem}

\begin{theorem}[consistency]
Focal parametricity is consistent.
\qed
\end{theorem}

We do not know if the term model of the parametric target calculus
satisfies the equalising requirement 
 --- if so, by the definability result, the parametricity on the 
target and the focal parametricity on $\lambda\mu2$ should agree. 
Alternatively we should consider a refined target calculus with
a construct ensuring the equalising requirement, as detailed in
Taylor's work on sober space (``a lambda calculus for sobriety'' \cite{Tay02}).
For now, we only know that one direction
is true (thanks to the definability).

\begin{theorem}
An equality derivable in $\lambda\mu2^P$ 
is also derivable in $\lambda\mu2$ with focal parametricity.
\qed
\end{theorem}

\section{Examples}

We show that certain impredicative encodings in $\lambda\mu2$
satisfy universal properties with respect to the focus
using the focal parametricity principle. 

\subsection{Focal decomposition}

We start with a remark on the following ``focal decomposition''
\cite{Sel01} (analogous to the linear decomposition
$\sigma_1\rightarrow\sigma_2=!\sigma_1\multimap\sigma_2$
\cite{Gir87}): there is a bijective correspondence between terms of
$\sigma_1\rightarrow\sigma_2$ and focal terms of
$\neg\neg\sigma_1\rightarrow\sigma_2$ natural in $\sigma_1$ and focal
$\sigma_2$.
$$
\infer{f^\flat=f\circ \eta_{\sigma_1}=\lambda x^{\sigma_1}.f\,(\lambda k.k\,x):\sigma_1\rightarrow\sigma_2}
      {f:\neg\neg \sigma_1\rightarrow\sigma_2~~\mathrm{focal}}
$$

$$
\infer{
g^\sharp=C_{\sigma_2}\circ\neg\neg g=
 \lambda m.\mymu\beta^{\sigma_2}.m\,(\lambda x^{\sigma_1}.\myname{\beta}(g\,x))
:\neg\neg \sigma_1\rightarrow\sigma_2~~\mathrm{focal}
}
      {g:\sigma_1\rightarrow\sigma_2}
$$
\begin{proposition}
${g^\sharp}^\flat=g$ for any $g:\sigma_1\rightarrow\sigma_2$,
while ${f^\flat}^\sharp=f$ holds for $f:\neg\neg \sigma_1\rightarrow\sigma_2$
if and only if $f$ is focal.
\qed
\end{proposition}

\subsection{Falsity as a focally initial object}

Now we shall proceed to reason about impredicative encodings in $\lambda\mu2$.
The first example is the falsity $\bot=\forall X.X$.

First, we note that 
$A_\sigma=\lambda x^\bot.x\,\sigma:\bot\rightarrow\sigma$ is focal.
The parametricity on $\bot$ says $x\,\bot^*\,x$ for any $x:\bot$.
Since $\pair{A_\sigma}:\bot\leftrightarrow\sigma$, we have
$x\,\bot\,\pair{A_\sigma}\,x\,\sigma$, i.e. 
$A_\sigma\,(x\,\bot)=x\,\bot\,\sigma=x\,\sigma$. By extensionality we
get $x=x\,\bot$ for $x:\bot$. 

Now suppose that $g:\bot\rightarrow\sigma$ is focal.
Again by the parametricity on $\bot$ we know $x\,\bot^*\,x$ for any $x:\bot$,
hence $x\,\bot\,\pair{g}\,x\,\sigma$. Thus $g\,(x\,\bot)=x\,\sigma$;
but $x=x\,\bot$, 
so we have $g\,x=x\,\sigma$, hence $g=\lambda x^\bot.x\,\sigma=A_\sigma$.

So we conclude that $A_\sigma$ is the 
unique focal map from $\bot$ to $\sigma$. This means that
$\bot$ is initial in the focus.

\subsection{Focally initial algebra}

As in $\lambda2$, there is a fairly standard encoding
$$
\begin{array}{rcl}
\mu X.F[X] &=& \forall X.(F[X]\rightarrow X)\rightarrow X\\
\mathsf{fold}_\sigma
 &=& \lambda a^{F[\sigma]\rightarrow\sigma}.
     \lambda x^{\mu X.F[X]}.x\,\sigma\,a
 :(F[\sigma]\rightarrow\sigma)\rightarrow\mu X.F[X]\rightarrow\sigma \\
\mathsf{in} &=& \lambda y.\Lambda X.\lambda k^{F[X]\rightarrow X}.k\,(F[\mathsf{fold}_X\,k]\,y)
 :F[\mu X.F[X]]\rightarrow\mu X.F[X]\\
\end{array}
$$
for which the following diagram commutes (just by $\beta$-axioms).
$$
\begin{diagram}
\node{F[\mu X.F[X]]}
 \arrow{e,t}{\mathsf{in}}
 \arrow{s,l}{F[\mathsf{fold}\,a]}
\node{\mu X.F[X]}
 \arrow{s,r}{\mathsf{fold}\,a}\\
\node{F[\sigma]}
 \arrow{e,b}{a}
\node{\sigma}
\end{diagram}
$$
Therefore $\mathsf{in}$ is a {\em weak} initial $F$-algebra.
However, as we noted before, $\mathsf{in}$ is not an initial $F$-algebra ---
in fact it is not even an isomorphism. 
By applying the focal decomposition above, we obtain the
commutative diagram
$$
\begin{diagram}
\node{\neg\neg F[\mu X.F[X]]}
 \arrow{e,t}{\mathsf{in^\sharp}}
 \arrow{s,l}{\neg\neg F[\mathsf{fold}\,a^\flat]}
\node{\mu X.F[X]}
 \arrow{s,r}{\mathsf{fold}\,a^\flat}\\
\node{\neg\neg F[\sigma]}
 \arrow{e,b}{a}
\node{\sigma}
\end{diagram}
$$
for any focal $a:\neg\neg F[\sigma]\rightarrow \sigma$.
We show that $\mathsf{fold}\,a^\flat$ is the unique {\em focal} map
making this diagram commute, thus $\mathsf{in}^\sharp$ 
is an initial $\neg\neg F[-]$-algebra in the focus.

We sketch a proof which is fairly analogous to that for the
corresponding result in parametric $\lambda2$ as given in \cite{ACC93}.
First, from the parametricity on $\mu X.F[X]$ we obtain that
$$
\begin{diagram}
\node{F[\sigma_1]}
 \arrow{e,t}{a}
 \arrow{s,l}{F[h]}
\node{\sigma_1}
 \arrow{s,r}{h}\\
\node{F[\sigma_2]}
 \arrow{e,b}{b}
\node{\sigma_2}
\end{diagram}
~~~\mbox{implies}
\begin{diagram}
\node{\mu X.F[X]}
 \arrow{e,t}{\mathsf{fold}\,a}
 \arrow{s,l}{||}
\node{\sigma_1}
 \arrow{s,r}{h}\\
\node{\mu X.F[X]}
 \arrow{e,b}{\mathsf{fold}\,b}
\node{\sigma_2}
\end{diagram}
$$
whenever $h$ is focal. We also have
$M\,(\mu X.F[X])\,\mathsf{in}=M$ for any $M:\mu X.F[X]$ as
a corollary (thanks to extensionality). 
By combining these observations, now we have the desired result.
That is, if $h:\mu X.F[X]\rightarrow\sigma$ is focal and satisfies
$h\circ\mathsf{in}^\sharp=a\circ\neg\neg F[h]$, then
$$
\begin{array}{rcl}
\mathsf{fold}_\sigma\,a^\flat\,x
&=&
h\,(\mathsf{fold}_{\mu X.F[X]}\,\mathsf{in}\,x)\\
&=&
h\,(x\,(\mu X.F[X])\,\mathsf{in})\\
&=&
h\,x
\end{array}
$$
so by extensionality we conclude $\mathsf{fold}_\sigma\,a^\flat=h$.
This also implies that 
$\mathsf{in}^\sharp$ is an isomorphism,
with the inverse given by
$\mathsf{fold}_{\neg\neg\mu X.F[X]}\,(\neg\neg F[\mathsf{in}])$.


As a special case, 
by letting $F$ be a constant functor,
we obtain isomorphisms between
$(\sigma\rightarrow\bot)\rightarrow\bot$ and
$\forall X.(\sigma\rightarrow X)\rightarrow X$ 
where $X$ is not free in $\sigma$.
With some further calculation we see that 
$
\mathsf{in}^\sharp=
\lambda m.\Lambda X.\lambda k^{\sigma\rightarrow X}.\mymu\alpha^X.m\,(\lambda x^\sigma.\myname{\alpha}(k\,x))$
is the inverse of
$\lambda n.n\,\bot:(\forall X.(\sigma\rightarrow X)\rightarrow X)\rightarrow
 (\sigma\rightarrow\bot)\rightarrow\bot$.
We will see more about this isomorphism in Section 7.

\subsection{The type of Church numerals}

\newcommand{\zero}{\mathsf{O}}
\newcommand{\suc}{\mathsf{S}}
\newcommand{\bfN}{\mathbf{N}}
\newcommand{\F}[1]{\bot\rightarrow(#1\rightarrow\bot)\rightarrow\bot}

We conclude this section by a remark on the type of Church numerals
$\mathbf{N}=\forall X.X\rightarrow(X\rightarrow X)\rightarrow X$.
Recall that, in $\lambda2$ with parametricity, $\mathbf{N}$ is an initial algebra
of $\forall X.X\rightarrow(-\rightarrow X)\rightarrow X
 \simeq 1+(-)$, i.e. a natural numbers object, whose closed 
inhabitants are equal to the Church numerals $\suc^n\,\zero$
which can be given by, as usual,
$$
\begin{array}{lclcl}
\zero &=& \Lambda X.\lambda x^X f^{X\rightarrow X}.x &:& \mathbf{N}\\
\suc &=& 
 \lambda n^\bfN.\Lambda X.\lambda x^X f^{X\rightarrow X}.f\,(n\,X\,x\,f) 
 &:& \bfN\rightarrow\bfN
\end{array}
$$
It is no longer true in $\lambda\mu2$, as observed by Parigot, 
as there are closed inhabitants which are not equal to Church numerals, e.g.
$$
 \mu\alpha^\mathbf{N}.[\alpha](\suc\,(\mu\beta^\mathbf{N}.[\alpha]\zero))=
 \Lambda X.\lambda x^X f^{X\rightarrow X}.\mu\alpha^X.[\alpha]
 (f\,(\mu\beta^X.[\alpha]x)):\mathbf{N}$$
In contrast, $\mathbf{N}$ in $\lambda\mu2$ with focal parametricity 
is a focally initial algebra of 
$\forall X.X\rightarrow(-\rightarrow X)\rightarrow X
 \simeq
 \bot\rightarrow(-\rightarrow\bot)\rightarrow\bot$;
this can be shown in the same way as the case of focally initial algebras.
Spelling this out, we have a focal map
$\mathsf{in}:(\F{\mathbf{N}})\rightarrow\mathbf{N}$,
and for any focal 
$g:(\F{\sigma})\rightarrow \sigma$
there exists a unique focal $\mathsf{fold}_\sigma\,g:\mathbf{N}\rightarrow \sigma$
making the following diagram commute.
$$
\begin{diagram}
\node{\F{\mathbf{N}}}
 \arrow{e,t}{\mathsf{in}}
 \arrow{s,l}{\F{\mathsf{fold}_\sigma\,g}}
\node{\mathbf{N}}
 \arrow{s,r}{\mathsf{fold}_\sigma\,g}
\\
\node{\F{\sigma}}
 \arrow{e,b}{g}
\node{\sigma}
\end{diagram}
$$
To see this, it is useful to observe the following 
bijective correspondence  (a variant of the focal decomposition):
given focal $g:(\F{\sigma})\rightarrow \sigma$ we have
$$
\begin{array}{lclcl}
g_o &=& g\,(\lambda x^\bot k^{\sigma\rightarrow\bot}.x) &:& \sigma\\
g_s  &=& \lambda y^\sigma.g\,(\lambda x^\bot k^{\sigma\rightarrow\bot}.k\,y) &:&
 \sigma\rightarrow \sigma
\end{array}
$$
and conversely, for $a:\sigma$ and $f:\sigma\rightarrow\sigma$ we have a focal map
$$
\varphi_{a,f} ~=~
 \lambda m^{\F{\sigma}}.\mu\alpha^\sigma.m\,([\alpha]a)\,(\lambda y^\sigma.[\alpha](f\,y))
~:~(\F{\sigma})\rightarrow \sigma
$$
It follows that $(\varphi_{a,f})_o=a$ and
$(\varphi_{a,f})_s=f$ hold for any $a$ and $f$,
while 
$\varphi_{g_o,g_s}=g$ for any focal $g$.
Now we define
$$
\begin{array}{lclcl}
\mathsf{fold}_A\,g &=& \lambda n^\bfN.n\,A\,g_o\,g_s
  &:& \bfN\rightarrow A\\
\mathsf{in} &=& \varphi_{\zero,\suc}
  &:& (\F{\bfN})\rightarrow\bfN
\end{array}
$$
It then follows that the diagram above commutes --- and the focal 
parametricity implies that $\mathsf{fold}_A\,g$ is the unique 
such focal map.

\section{A general characterisation}

So far, we concentrated on the relational parametricity for
$\lambda\mu2$. One may feel that this story is very specific
to the case of $\lambda\mu2$, or of the first-class continuations, 
and is not immediately applicable to other computational effects.

In this section we describe an alternative characterisation of
the focus, which makes sense in any extension of $\lambda2$. 
Namely, we show that, any $\lambda2$-theory is equipped with a
monad $L$, such that each type is equipped with an algebra structure
--- and then see that, in the case of $\lambda\mu2$ with focal
parametricity, this monad $L$ is isomorphic to the double-negation
(continuation) monad, and focal maps are precisely the
algebra maps of the monad $L$. 
This suggests a
natural generalisation of this work to a theory of parametricity 
for general computational effects.

\subsection{A monad on \protect{$\lambda2$}}
\newcommand{\poly}[1]{\forall X.(#1\rightarrow X)\rightarrow X}

Let $L\sigma=\poly{\sigma}$ (with no free $X$ in $\sigma$), and define
$$
\begin{array}{lllll}
\eta_\sigma &=& \lambda x^\sigma.\Lambda X.\lambda k^{\sigma\rightarrow X}.k\,x
       &:& \sigma\rightarrow L \sigma\\
\mu_\sigma  &=& \lambda z^{L^2\sigma}.\Lambda X.\lambda k^{\sigma\rightarrow X}.z\,X\,(\lambda y^{L\sigma}.y\,X\,k)
       &:& L^2\sigma\rightarrow L\sigma\\
L(f)   &=& \lambda y^{L\sigma_1}.\Lambda X.\lambda h^{\sigma_2\rightarrow X}.y\,X\,(h\circ f)
       &:& L\sigma_1\rightarrow L\sigma_2~~(f:\sigma_1\rightarrow \sigma_2)\\
\end{array}
$$     
\begin{proposition}
On the term model of any $\lambda2$-theory, $(L,\eta,\mu)$ forms a monad.
\qed
\end{proposition}
One might think that this is trivial as
$L\sigma$ is isomorphic to $\sigma$ when we assume the 
standard parametricity. This is not always the case however, as we have 
already seen, $L\sigma\simeq (\sigma\rightarrow\bot)\rightarrow\bot$ in 
the focally parametric $\lambda\mu2$.
\begin{proposition}
$\alpha_\sigma=\lambda y^{L\sigma}.y\,\sigma\,(\lambda x^\sigma.x):
L\sigma\rightarrow\sigma$ is an algebra of the monad $(L,\eta,\mu)$.
\qed
\end{proposition}
Thus each $\sigma$ is canonically equipped with 
an algebra structure $\alpha_\sigma$.
Again one may think that this is trivial, as
under the standard parametricity $\alpha_\sigma$ is just an isomorphism 
with $\eta_\sigma$ being an inverse. However, again it is not the case in
a non-trivial $\lambda\mu2$-theory.

Now we define the notion of linear maps in terms of the monad $L$ and
the canonical algebras $\alpha_\sigma$ --- this is close to what we do
in (axiomatic) domain theory for characterising  the strict maps, and also
in control categories for characterising the focal maps.

\begin{definition}
$f:\sigma_1\rightarrow\sigma_2$ is {\em linear} when 
it is an algebra morphism from $\alpha_{\sigma_1}$ to $\alpha_{\sigma_2}$, i.e.
$f\circ\alpha_{\sigma_1}=\alpha_{\sigma_2}\circ L(f)$ holds. 
\end{definition}
$$
\begin{diagram}
\node{L\sigma_1}
 \arrow{e,t}{L(f)}
 \arrow{s,l}{\alpha_{\sigma_1}}
\node{L\sigma_2}
 \arrow{s,r}{\alpha_{\sigma_2}}\\
\node{\sigma_1}
 \arrow{e,b}{f}
\node{\sigma_2}
\end{diagram}
$$
That is, 
$f$ is linear when
$$
f\,(M\,\sigma_1\,(\lambda x^{\sigma_1}.x))=M\,\sigma_2\,f
$$
holds for any $M:L\sigma_1$. We may write $f:\sigma_1\multimap\sigma_2$
for a linear $f:\sigma_1\rightarrow\sigma_2$. 
Under the standard parametricity every $f:\sigma_1\rightarrow\sigma_2$ is 
linear, while for focal parametricity on $\lambda\mu2$ we have that
linear maps are precisely the focal maps (see below).
In passing, we note the following interesting observation.

\begin{proposition}
In a $\lambda2$-theory, the following conditions are equivalent.
\begin{enumerate}
\item algebras on $\sigma_1\rightarrow\sigma_2$ and $\forall X.\sigma$ are 
determined in the pointwise manner, i.e.
$$
\begin{array}{rcl}
\alpha_{\sigma_1\rightarrow \sigma_2}
&=&
\lambda f^{L(\sigma_1\rightarrow\sigma_2)}.\lambda x^{\sigma_1}.\alpha_{\sigma_2}\,(L(\lambda g^{\sigma_1\rightarrow \sigma_2}.g\,x)\,f)\\
\alpha_{\forall X.\sigma}
&=&
\lambda x^{L(\forall X.\sigma)}.\Lambda X.\alpha_\sigma\,(L(\lambda y^{\forall X.\sigma}.y\,X)\,x)
\end{array}
$$
\item $\lambda x^{\sigma_1\rightarrow\sigma_2}.x\,N$ is linear
  for any $N:\sigma_1$, and
  $\lambda x^{\forall X.\sigma}.x\,\sigma_1$ is linear
  for any $\sigma$ and $\sigma_1$.\qed
\end{enumerate}
\end{proposition}
Note that they are very close to the ``additional axioms'' for 
$\lambda\mu2$ discussed
in Section 5. Also note that, if a $\lambda2$-theory satisfies one of these 
conditions, $\mu_\sigma$ and $\alpha_{L\sigma}$ agree for every $\sigma$.
and
we have a ``linear decomposition'' correspondence
between the maps of $\sigma_1\rightarrow \sigma_2$ and 
the linear maps of $L\sigma_1\rightarrow\sigma_2$.

\subsection{Focal maps as algebra maps}

Now we shall consider the double-negation monad
$\neg\neg\sigma=(\sigma\rightarrow\bot)\rightarrow\bot$ on 
$\lambda\mu2$ with focal parametricity.

\begin{proposition}
In a focally parametric $\lambda\mu2$-theory,
$C_\sigma:\neg\neg\sigma\rightarrow\sigma$ is an algebra of the
double-negation monad.
\qed
\end{proposition}

\begin{corollary}
$f:\sigma_1\rightarrow\sigma_2$ is focal if and only if it is 
an algebra map from $C_{\sigma_1}$ to $C_{\sigma_2}$.
\qed
\end{corollary}

\begin{proposition}
The monad $(L,\eta,\mu)$ is isomorphic to the double negation monad
in the focally parametric $\lambda\mu2$, with
$\lambda x^{L\sigma}.x\,\bot:
 L\sigma\stackrel{\simeq}{\rightarrow}\neg\neg\sigma$.
\qed
\end{proposition}

\begin{proposition}
The following diagram commutes in a focally parametric $\lambda\mu2$-theory:
$$\hbox to 159.6pt{\hfil}
\begin{diagram}
\node{L\sigma} \arrow[2]{e,t}{\lambda x^{L\sigma}.x\,\bot}
\arrow{se,r}{\alpha_\sigma}
\node[2]{\neg\neg\sigma}
\arrow{sw,r}{C_\sigma}
\\
\node[2]{\sigma}
\end{diagram}\hbox to 159.6pt{\hfil}\qED{20}
$$
\end{proposition}

\begin{corollary}
$f:\sigma_1\rightarrow\sigma_2$ is linear if and only if it is focal.
\qed
\end{corollary}
Thus a focal map in a focally parametric $\lambda\mu2$-theory can be
characterised just in terms of the monad $L$ which is defined for
arbitrary $\lambda2$-theory. 

We believe that the monad $L$ deserves much attention. It has been 
considered trivial, but now we know that it does characterise an
essential notion (focus) in the case of relational parametricity 
under the presence of control feature. In fact, the story does not
end here; under the presence of non-termination or recursion,
$L$ behaves like a {\em lifting monad} --- indeed it is a
lifting in the theory of linear parametricity, because
$$
L\sigma
~=~
\forall X.(\sigma\rightarrow X)\rightarrow X
~=~
\forall X.!(!\sigma\multimap X)\multimap X
~\simeq~
!\sigma
$$
where the last isomorphism follows from the fact that 
$\forall X.!(F[X]\multimap X)\multimap X$ gives an
initial algebra of $F$, cf. \cite{BPR00}.

These observations suggest that there exists a general framework 
similar to (axiomatic or synthetic) domain theory where the lifting
monad can be replaced by any strong monad --- a continuation monad
for example --- on which a theory of 
parametricity for general computational effects can be built.
Recently, Alex Simpson has made a progress in this direction, by
developing a two-level polymorphic type theory 
(for interpreting types and algebras of a monad) in a 
constructive universe \cite{Sim05}. His work fits very well with 
the case of linear parametricity for recursion; it is plausible
that it also explains the case of focal parametricity 
for first-class control.

\section{Conclusion and future work}

We have studied the relational parametricity for $\lambda\mu2$,
first by considering the CPS translation into a parametric fragment 
of $\lambda2$, and then by directly giving a constrained parametricity
for $\lambda\mu2$. The later, which we call ``focal parametricity'',
seems to be a natural parametricity principle under the presence of
first-class controls --- in the same sense that linear 
parametricity works under the presence of recursion and non-termination.

There remain many things to be addressed in future. 
In the previous section, we already discussed a research direction 
towards a relational parametricity for general effects.
Below we shall briefly mention some future work more closely 
related to the main development of this paper. 

Firstly, we are yet to complete the precise comparison between 
focal parametricity on $\lambda\mu2$ and the parametricity on the 
CPS target calculus. This involves some subtle interaction between
parametricity and a technical condition (equalising requirement).

Secondly, we should study focal parametricity for  
extensions of $\lambda\mu2$. As we observed, $\lambda\mu2$
with focal parametricity does not have many popular datatypes,
e.g. cartesian products, and classical disjunction types 
which however can be added with no problem. Adding general
initial algebras is problematic (having an initial object already means 
inconsistency), but it might be safe to add certain carefully chosen 
instances. On the other hand, final coalgebras seem less 
problematic, though a generic account for them in $\lambda\mu2$ is
still missing. Perhaps we also need to consider 
the CPS translation of such datatypes (cf. \cite{BU02}) in a systematic
way.

An interesting topic we have not discussed in this paper
is the Filinski-Selinger duality \cite{Fil89,Sel01} between
call-by-name and call-by-value calculi with control primitives. 
In fact it is straightforward to consider its second-order extension:
in short, universal quantifiers in call-by-name (as studied in this paper)
amount to existential quantifiers in call-by-value. 
We are not sure if the call-by-value calculus with existential quantifiers 
itself is of some interest. However, it can be a good starting point to
understand the call-by-value parametric polymorphism (possibly with 
computational effects), from both syntactic and semantic aspects.
In particular,
it should provide new insights on the famous difficulty of accommodating
first-class continuations in ML type system \cite{HDM93}. 

Finally, we also should consider if there is a better (ideally
semantic) formulation of focal relations. In this paper we only
consider those coming from focal maps, but it seems natural to regard
a subalgebra (of the double-negation monad) of
$C_{\sigma_1\times\sigma_2}$ as a focal relation between $\sigma_1$
and $\sigma_2$, where we assume the presence of cartesian product
$\sigma_1\times\sigma_2$. This looks very closely related to Pitts'
$\top\top$-closed relations for $\lambda2$ with recursion
\cite{Pit00}.

\section*{Acknowledgement}
I thank Ken-etsu Fujita for discussions and cooperations related to
this work.
I am also grateful to Ryu Hasegawa, Paul-Andr\'e Melli\`es and 
Alex Simpson for comments, discussions and encouragements.

\end{document}